\documentclass[twocolumn,aps,showpacs,preprintnumbers,nofootinbib,
superscriiptaddress]{revtex4}
\usepackage{graphicx}
\usepackage{dcolumn}
\usepackage{bm}
\begin{document}

\title{Energy losses and efficiency of laser -- electron X-ray
generator for medical applications}

\author{E.G.~Bessonov, R.M.Fechtchenko, M.V. Gorbunkov,
A.V. Vinogradov \\
{\it Lebedev Physical Institute RAS, Moscow, Russia} \\
V.I. Shvedunov         \\
{\it Nuclear Physics Institute of Moscow State University, 119899 Russia,
Moscow, Vorobyevy Gory}}

                       \begin{abstract}
A source of medical x-rays based on a 50 Mev storage ring and a
quasi-continues picosecond laser is considered. It is shown that such
generator produces useful X-ray flux with higher average power and higher
efficiency than that of conventional X-ray tubes. The main energy losses are
related to coherent synchrotron radiation. Taking this into account the
required parameters of the storage ring and injector are determined.
\end{abstract}


\maketitle

\vspace{-3mm}
\section{Introduction}

The experiments with synchrotron radiation sources confirmed that the
use of narrowband X-ray beams extends substantially diagnostics
capabilities in mammography, bronchography, angiography, computer
tomography etc \cite {1} - \cite {3}. Apart from the image contrast
improvement there are several other advantages of this method:
possibility for subtraction of images taken at the wavelengths before
and after absorption K-edge of the contrast element (usually iodine),
reduction of radiation dose and amount of the contrast material,
intravenous (catheterless) angiography of coronary arteries, etc.
Actually new attractive methods of X-ray diagnostics are emerging.

However further development and practical applications of these methods
depend on the availability of a compact (as compared with synchrotrons)
narrowband X-ray source suitable for installation in medical centers
and hospitals. As a rule it should work in the energy range 20 -50 kev
that includes K absorption edges of the elements I and Gd used in
contrast compounds.

As a solution of the problem of narrowband compact X-ray source it was
proposed to generate X-rays in collisions of a laser beam with an
electron beam moving in opposite direction. In such an approach the
electron beam energy decreases from several Gev (as for the systems
based on synchrotrons with magnetic undulators) to tens Mev and the
dimension of the device changes from tens of meters down to several
meters. To produce electron beams for this purpose both storage rings
\cite{4}, \cite{5} and linear accelerators \cite {6} can be utilized.
In this paper the energy losses and efficiency of X-ray generator based
on a storage ring and repetitive laser is considered (see FIG. 1).

\vspace{-3mm}
\section{Laser-electron X-ray generator (LEXG)}

The X-ray photon flux required for coronary arteries subtraction angiography
is estimated as:

\vspace{-3mm}
\begin{equation}
\label{eq1}
\Phi \approx 2 \cdot 10 ^{14} s ^{-1 }
\end{equation}

\noindent
at the surface of a human body \cite{7}, \cite {8}. The time necessary
to obtain a single image must be 2-4 ms, and the frame rate: 20 $\div$
30 s$^{ - 1}$ \cite {9}.

\vspace{1mm}

\begin{figure}[hbt]
\includegraphics{
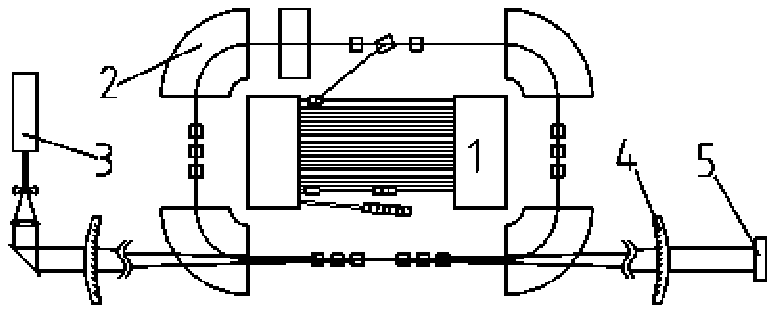}  
\caption{
The scheme of laser-electron X-ray generator: 1 - race-track microtron, 2 --
storage ring, 3 - laser, 4 -- optical cavity, 5 -- a damp for laser
beam.}
\end{figure}

\vspace{-1mm}

In the electron rest frame the process of laser photon scattering by an
electron is the classical Thompson scattering. The energy of the scattered
(X-ray) photon in the laboratory frame is:

\vspace{-3mm}
\begin{equation}
\label{eq2}
\hbar \omega = 4\gamma ^2\hbar \omega _L ,
\end{equation}

\noindent
where $\hbar \omega _L $ is laser photon energy and $\gamma = E / mc^2$,
$E_{ }$ electron energy. The average X-ray photon flux caused by
Thomson scattering is:

\begin{equation}
\label{eq3}
\Phi = N_e \frac{\sigma _T }{s}\Phi _L \left( {1 - r} \right)^{ - 1},
\hskip 8mm s = s_{e}+s_{L}, \end{equation}

\noindent
where $N_{e}$ is the number of electrons in the bunch; $\sigma _T = 6.6
\cdot 10^{ - 25}$cm$^{2}$, Thompson cross-section; $\Phi _{L}$, the
laser photon flux; $\left( {1 - r} \right)^{ - 1}$, the finesse of
the optical cavity, which is used to enhance the power of the laser
beam in the interaction region; $s_{e }$and $s_{L}$, the cross section
areas of the electron and laser beams respectively. We assume that both
laser and electron beams consist of bunches following with equal
repetition rate and having equal cross-section areas $s_{e}=s_{L}=s$/2
(see Appendix A).

Using (\ref{eq1}), (3) it is easy to show that X-ray flux required by
(1) can be attained with a system consisting of a storage ring, a laser
and an optical cavity, which parameters are given in Table I. The
listed parameters either have been already achieved or close to the
achieved ones in existing setups \cite{10}.

The electron bunch length in Table I was chosen equal to double Raleigh
length of laser beam: $l_e = 2z_L = 4s_L / \lambda = 4s_e / \lambda $
assuming the Gaussian electron density distribution in transverse direction
(see Appendix A). Then the storage ring emittance is $ \sim $0.1 mm$ \cdot
$mrad.

\vspace{5mm}
\underline {Table I. Parameters of the LEXG.}

\vspace{-0mm}
\begin {center}
\underline{Storage ring}
\end {center}

\vspace{-3mm}
\vbox {
\noindent
Electron energy \hfill $E=46$ MeV\\
Relative energy \hfill $\gamma = 91$ \\
Bending radius \hfill $R = 05$ m \\
Orbit circumference \hfill $C = 6$ m \\
Revolution period \hfill $T = 20$ ns \\
Number of electrons in bunch $\hfill 1.25 \cdot 10 ^{10}$\\
Current \hfill  $J =0.1$ A \\
Bunch length \hfill $l_{e} = 1.2$ cm \\

\vspace{-5mm}
\begin {center}
\underline{Laser}
\end {center}

\vspace{-4mm}
\noindent
Laser photon energy \hfill $\hbar \omega _L = 1$ eV \\
Repetition frequency \hfill $F = 50$ MHz \\
Laser beam cross section area \hfill $s _L =3 \cdot 10 ^5$ cm$^2$ \\
The laser pulse energy \hfill $\varepsilon _L = 1.2$ $\mu$m \\
The average laser power \hfill $P_{L}= F \cdot \varepsilon _L = 60$ W\\
The laser pulse duration \hfill $\tau _L < 40$ ps\\

\vspace{-5mm}
\begin {center}
\underline{Cavity}
\end {center}

\vspace{-0mm}
\noindent
Finesse $(1 - r) ^{-1} = 4 \cdot 10^3$ \\
The energy of the laser pulse in the cavity \hfill $\varepsilon _r =
4.8$ mJ\\

}

\vspace{3mm}
In the next sections we estimate the electron beam energy losses and the
X-ray generator efficiency.

\vspace{-3mm}
\section{The synchrotron radiation losses}

The main source of electron energy losses in our storage ring (see Table I)
is coherent synchrotron radiation (CSR), which spectrum lies in the
microwave region. The CSR power for a bunch with Gaussian longitudinal
charge density distribution can be estimated as (see Appendix A):

\vspace{-3mm}
\begin{equation}
\label{eq4}
P = \frac{3^{1 / 6}\Gamma ^2\left( {2 / 3} \right)r_e cN^2}{2\pi R^{2 /
3}l^{4 / 3}(1 + \mu )}mc^2
\end{equation}

\noindent
where $r_{e}$ = 2.8$ \cdot $10$^{ - 13 }$cm, $1 + \mu = C / 2\pi R, C $
is the perimeter of the electron orbit.

Substituting the values from Table I for the LESG into (\ref{eq2}) and also
supposing $\mu = 1$ we obtain:

\vspace{-3mm}
\begin{equation}
\label{eq5}
P =1.1 kW, \hskip 15mm \Delta E = 11 kev,
\end{equation}

\noindent
where $\Delta E$ is the energy lost by an electron in one revolution.
This energy loss is compensated by the storage ring radio frequency
(RF) cavity.  The accelerating voltage there should not be less than
\textit{U=250 kV} in order to provide the stability of the electron
bunch (see Appendix B). In the next section the storage ring power
consumption is estimated.

\vspace{-3mm}
\section{ Storage ring and injector}

As an injector for the storage ring it is reasonable to take a compact
highly efficient race-track microtron with photo-cathode electron gun
\cite {11}.  To simplify storage ring operation the injection energy
should be close to the final energy of the stored electrons. To store
$\sim $1 nC charge in a single ring bunch multi-turn injection can be
used with the injector operating in single bunch mode with 100-150 pC
bunch charge and bunch repetition frequency dependent on the ring
damping time. The use of rare-earth permanent magnets in the injector
and storage ring design will essentially decrease energy consumption.

In the storage ring with permanent magnets the required electric power is
defined mainly by an RF power $P_{RF}$ necessary to produce the accelerating
structure voltage $U$ inside the accelerating structure (single RF cavity or a
chain of coupled cavities):

\vspace{-3mm}
\begin{equation}
\label{eq6}
P_{RF} = \frac{U^2}{ZL},
\end{equation}

\noindent
where $Z$ is the effective shunt impedance and $L$ is the structure's
electrical length. Because of $Z \propto \sqrt {f_{RF} } $, where $f_{RF}$ is
operating frequency, for a fixed structure length the RF power required
decreases with frequency increase. Some aspects of storage ring beam
dynamics are also favorable to higher frequency. Other considerations, in
particular the separatrix size, the cavity beam hole aperture favors lower
frequency. The final choice hasn't been made yet, so two possibilities were
considered: 2856 MHz -- equal to the injector operating frequency
\cite {11}, and 571.2 MHz -- the fifth subharmonic. Consider elliptical
form cavity, which has 1.5-2 times lower shunt impedance as compared
with the optimized cavity, but is more stable with respect to
multipactor discharge and parasitic modes excitation. With such a
cavity one obtains $Z$~$ \approx $~30~M$\Omega $/m and $P_{RF}$~$
\approx $~8~kW for $L$ = 0.2625 m (five half wavelength coupled
cavities chain) at 2856 MHz, and $ Z \quad  \approx $ 13.4 M$\Omega
$/m, $P_{RF} \quad \approx $ 18 kW for the single cavity of the same
length at 571.2 MHz. The RF power dissipated per unit length in the
considered cases is 30 kW/m and 69 kW/m, respectively. These thermal
loads will not change essentially structure parameters if the cooling
channels are properly constructed.  Taking into account beam loading
and RF amplifier efficiency, the full power consumption from the socket
will be about 20 and 40 kW, respectively.

\vspace{ -8mm}
\section{The source of the optical radiation and its efficiency}

Power consumption of optical unit of LEXG can be an order of magnitude less
than that required for RF system if a solid state laser with diode pumping
is used. Such lasers reached tens of percent efficiency during last decade.
For example, in [12] the efficiency 28{\%} is reported for Nd:YAG laser
having quasi-continuous output with average power more than 300 W. The
active element was 6 mm diameter and 50 mm length cylinder pumped by six
high power diode bars.

Besides of efficiency, the stability of laser radiation is one more
important parameter for LEXG. The enhancement of laser light power inside
high finesse optical resonator implies both equal repetition rates and
careful phase matching of laser and resonator pulses. In other words the
most part of every laser mode energy must be concentrated inside the
bandwidth $\Delta \nu _r $ of the corresponding resonator mode,$\Delta \nu
_r = c\left( {1 - r} \right) / 2\pi \cdot r^{1 / 2}L_c $, where $L_{c}$ is
the length of the cavity. From this point of view optically pumped solid
state lasers generating subpicosecond- and femtosecond pulses again have
evident advantages [13]. For example, crystalline active mediums doped with
ytterbium are successfully utilized for continues sequences of powerful
subpicosecond pulses production. In [14] the average power of 60 W was
obtained in a Yb:YAG laser with the duration of a single pulse 0.81 ps, and
optical power of a linier diode assembly being 370 W. Estimated efficiency
is not less than 5{\%}.

A significant decrease of the optical source average power can be achieved
with the transition from quasi-continues ultrashort pulse sequence to
generation of pulse trains with train duration 2-4 ms corresponding to
accumulation time of one X-ray image and with train repetition rate being
20-30 Hz. Such mode of operation can be reached using external negative
feed-back loop \cite {10}.

Thus from the above discussion it follows that the power necessary to feed
the optical unit of LEXG is less than 2-3 kW.

\vspace{-3mm}
\section{ Discussion}

Estimations given in sections 4 and 5 show that for production of $\sim $1 W
(see (1)) of X-ray radiation with photon energy $\hbar \omega _L = 33$ kev a
laser-electron X-ray generator with electrical power consumption 25-45 kW is
required. In respect of efficiency (the ratio of useful X-ray flux to
electric power consumption) LEXG surpasses X-ray tubes. Note that medical
X-ray tubes utilized in angiography have power up to 100 kW
\cite{9}, \cite{15}, \cite{16}.  No doubts that if physical and
technical problems connected with the development of LEXG are resolved
such X-ray sources will find wide application in medicine and other
fields.

Except the scheme shown in figure 1 there are other possibilities to
increase the coupling of electron and laser beams in order to provide
necessary X-ray flux. For instance high-energy laser pulse circulation
obtained by switching of a Pockel's cell inside the optical resonator may be
used [17]. The practical optical scheme should satisfy several conditions:
resistance of the mirrors and other optical elements to laser and X-ray
radiation loading, synchronization of laser pulses with electron bunches,
time and phase matching of optical resonator with the laser beam etc. This
will require special research.

\vspace{-3mm}
\section{ Conclusion}

A new X-ray generator for medical applications can be designed on the basis
of a storage ring with electron energy $\sim $50 Mev and a picosecond laser.
If high finesse optical resonator is used the plug efficiency of such a
generator can be higher than that of conventional X-ray tubes. Other
advantages of LEXG as compared to X-ray tubes are: narrowband X-ray beam,
the absence of high energy ``tail'' in photon spectrum, the possibility of
fine tuning of X-ray photons energy by changing the electron energy, image
quality improvement by subtraction of images taken at different wavelengths
near iodine K absorption edge, reduction of the necessary quantity of
contrast materials and radiation dose received by a patient and medical
staff, opportunity of catheterless angiography introduction into the wide
medical practice.

Thus, a compact laser-electron X-ray generator can be considered as a
promising direction of accelerator, laser and optical technology
development.

\vspace{-3mm}
         \section*{Appendix A}

Derivation of formula (3) assumed Gaussian radial density distribution of
electrons and photons:

\begin {equation}
\label{7}
\rho (x) = \frac{1}{\sqrt {2\pi } \sigma }e^{ - x^2 / 2\sigma ^2},
\end{equation}

\noindent
where $\sigma $ is the mean square transversal beam size connected with the
effective area of the cross section as $s = 2\pi \sigma ^2$. In the
longitudinal direction the mean square transversal laser beam size changes
in accordance with formula:

\vspace{-3mm}
\begin {equation}
\label{8}
\sigma _L (z) = \sigma _L \sqrt {1 + \frac{z^2}{z_L^2 }} ,
\end{equation}

\noindent
where $z_L = 2s_L / \lambda = 4\pi \sigma _L^2 / \lambda $ is the Raleigh
length.

The mean square transversal size of the electron beam is governed by the
same law. The role of Raleigh length in this case plays the $\beta
$-function at the central point of interaction of laser and electron beams.
The focal beam size is determined by the expression $\sigma _e = \left(
{\varepsilon \beta } \right)^{1 / 2}$, where $\varepsilon $ is the emittance
of the electron beam.

Formula (\ref{eq2}) for coherent synchrotron radiation losses was obtained by
L.Schiff in 1946 \cite {18}. Similar formulas were derived in
\cite{19} - \cite {21}.  However numerical coefficients there are
respectively 2$^{1 / 6} \quad \approx $1.12, 27$^{ / 3} \quad  \approx
$5.04 and 2$^{8 / 3} \quad \approx $6.35 times more than in
(\ref{eq2}). The analysis made by E.G. Bessonov and R.M. Fechtchenko
\cite {22} supported the results of L.Schiff.

\vspace{-3mm}
\section*{Appendix B}

Balance of energy gained by an electron in the RF cavity and lost due to
synchrotron radiation and Thomson scattering defines an equation for
electron phase oscillations in the storage ring \cite{23}, \cite {24}:

\begin {equation}
\label{9}
\frac{d^2\varphi }{dt^2} + \frac{k_i P_{noncoh}^{rad} }{E}\frac{d\varphi
}{dt} - \frac{he\Omega ^2}{2\pi E}\frac{\alpha \gamma ^2 - 1}{\gamma ^2 -
1}V\left( \varphi \right) = 0,
\end{equation}

\begin {equation}
\label{10}
V\left( \varphi \right) = \frac{2\pi }{\Omega e}\left[ {P_{coh}^{rad} \left(
\varphi \right) - P_{coh}^{rad} \left( {\varphi _s } \right)} \right] +
U\left( {\cos \varphi - \cos \varphi _s } \right),
\end{equation}

\noindent
where k$_{i}$ is the exponent in the energy dependence of incoherent
losses;  $P_{noncoh}^{rad} $, the total energy losses due to incoherent
synchrotron radiation and Thomson scattering; $P_{coh}^{rad} (\varphi
)$, the energy losses for coherent synchrotron radiation; $\Omega =
2\pi R / \left( {1 + \mu } \right)c$; the revolution frequency (which
is supposed to be equal to the laser repetition rate); $\alpha $, the
momentum compaction factor; $\varphi _s $, the particle phase in the
center of the bunch; the rest parameters were defined above. Equations
(9), (10) determine conditions for bunch stability in the storage
ring.

Incoherent synchrotron radiation and Thomson scattering provide slow damping
of the phase oscillations and can be neglected in the first approximation.
Coherent synchrotron radiation force acting on the electrons enlarges the
bunch length [20]. The stability criteria in terms of equation (9) are
provided by relation ${dV(\varphi )}/{d\varphi } < 0$ kept in the
vicinity of phase $\varphi _s $, occupied by bunch electrons. Using equation
for coherent synchrotron radiation power $P_{coh}^{rad} (\varphi )$
\cite {20} we obtain:

\begin {equation}
\label{11}
U > U_m = \frac{3^{1 / 6}2^{2 / 3}\left( {1 + \mu } \right)\Gamma ^2\left(
{2 / 3} \right)}{\sqrt {e_n } \sin \varphi _s }\frac{NeR^{4 / 3}}{hl_e^{7 /
3} },
\end{equation}
\vspace{-6mm}
$$\;\;\;e_n = 2,718... $$

Under this condition electron trajectories stay stable in spite of
destabilizing role of coherent synchrotron radiation.

Substituting $h$ =10, $\sin \varphi _s \approx 1$ and other parameters from the
Table I into (11) we get $U$ $ > $120~kV. In order the coherent
synchrotron radiation would not change essentially equilibrium bunch
dimensions this inequality must be fulfilled with some spare. More
details for electron beam dynamics in the coherent synchrotron
radiation field can be found in \cite{22}.

\end{document}